\documentclass [twocolumn, amsmath,amssymb,longbibliography,superscriptaddress]{revtex4-2}
\setcitestyle{numbers,square}
\usepackage{appendix}
\usepackage{array}
\usepackage{amsmath,amssymb}
\usepackage{tabularx}
\usepackage{graphicx}% Include figure files \usepackage{dcolumn}% Align table columns on decimal point
\usepackage{bmpsize}
\usepackage{bm}% bold math
\usepackage{color}
\usepackage{amssymb}
\usepackage [version=3]{mhchem}
\usepackage{newtxmath}
\usepackage{hyperref}
\usepackage{url}
\usepackage{physics}
\usepackage{here}
\usepackage{comment}
\DeclareMathAlphabet{\mathpzc}{OT1}{pzc}{m}{it}

\renewcommand{\vec}{\vb*}

\newcommand{\blue} [1]{\textcolor{blue}{ #1}}

\begin{document}
%\preprint{APS/123-QED}
%%%%%%%%%%%%%%%%%%%%%%%%%%%%%%%%%%%%%%%%%%%%%%%%%%%%%%%%%%%%%%%%%%%%%%%%%%%%%%%%%%%%%%%%%%%%%%%%%%%%%%%%%%%%%%%%%%%%%%%
\title{
Second-harmonic generation in twisted double bilayer graphene: Double-resonant enhancement from moir\'{e} flat bands
}

%\title{
%second-harmonic generation in twisted double bilayer graphene
%}

\author{Takaaki V. Joya}
\affiliation{Department of Physics, The University of Osaka, Toyonaka, Osaka 560-0043, Japan}

\author{Takuto Kawakami}
\affiliation{Center for Integrated Science and Humanities, Fukushima Medical University, Fukushima 960-1295, Japan}

\author{Mikito Koshino}
\affiliation{Institute for Solid State Physics, The University of Tokyo, Kashiwa, Chiba 277-8581, Japan}
\date{\today}
%%%%%%%%%%%%%%%%%%%%%%%%%%%%%%%%%%%%%%%%%%%%%%%%%%%%%%%%%%%%%%%%%%%%%%%%%%%%%%%%%%%%%%%%%%%%%%%%%%%%%%%%%%%%%%%%%%%%%%%

%%%%%%%%%%%%%%%%%%%%%%%%%%%%%%%%%%%%%%%%%%%%%%%%%%%%%%%%%%%%%%%%%%%%%%%%%%%%%%%%%%%%%%%%%%%%%%%%%%%%%%%%%%%%%%%%%%%%%%%
\begin{abstract}
We theoretically investigate second-harmonic generation (SHG) in twisted double bilayer graphene (TDBG) with AB--AB and AB--BA stacking configurations using a perturbative approach based on an effective continuum Hamiltonian.
We present a systematic analysis of the SHG response as a function of twist angle, vertical bias voltage, Fermi energy, and stacking configuration.
We find that the SHG signal is strongly enhanced at small twist angles due to the emergence of moir\'{e} flat bands and the associated increase in the joint density of states.
Beyond this conventional enhancement mechanism, we demonstrate that the reduced bandwidth enables a pronounced double-resonant process, in which optical transitions at both $\omega$ and $2\omega$ are simultaneously satisfied over extended regions of the moir\'{e} Brillouin zone.
This mechanism leads to a substantial amplification of the SHG response, analogous to the enhancement observed in systems with discrete energy levels, but realised here in a tuneable moir\'{e} band structure.
Furthermore, we show that the AB--AB and AB--BA configurations exhibit a systematic $\pi$ phase shift in the SHG response at large bias voltages, reflecting their distinct symmetry and electronic structure.
Our results identify double-resonance effects as a generic mechanism for enhancing SHG in moir\'{e} systems and provide a unified framework for understanding and controlling nonlinear optical responses in tuneable flat-band materials.
%We theoretically investigate second-harmonic generation (SHG) in AB--AB and AB--BA stacked twisted double bilayer graphene (TDBG) using a perturbative approach applied to an effective continuum Hamiltonian.
%We systematically analyse the dependence of the SHG spectrum on twist angle, vertical bias voltage, Fermi energy, and stacking configuration.
%We find that the signal is strongly enhanced at small twist angles due to the increased joint density of states associated with the formation of moir\'{e} flat bands.
%In particular, the response is further amplified by a double-resonance mechanism enabled by the tuneable flat bands.
%At large bias voltages, the two stacking configurations exhibit a systematic $\pi$ phase shift originating from their distinct geometries.

\end{abstract}
%%%%%%%%%%%%%%%%%%%%%%%%%%%%%%%%%%%%%%%%%%%%%%%%%%%%%%%%%%%%%%%%%%%%%%%%%%%%%%%%%%%%%%%%%%%%%%%%%%%%%%%%%%%%%%%%%%%%%%%

\maketitle
%%%%%%%%%%%%%%%%%%%%%%%%%%%%%%%%%%%%%%%%%%%%%%%%%%%%%%%%%%%%%%%%%%%%%%%%%%%%%%%%%%%%%%%%%%%%%%%%%%%%%%%%%%%%%%%%%%%%%%%
\section{Introduction}\label{sec:intro}
Second-harmonic generation (SHG) is a fundamental nonlinear optical process in which two photons of frequency $\omega$ are converted into a photon at frequency $2\omega$~\cite{FrankenWeinreich1961,Kleinman1962,Sipe2000,ParkerMorimoto2019}.
As a second-order process, it is only allowed in inversion-broken systems, and geometric quantities such as the Berry connection and quantum metric can play an important role in resonant responses~\cite{Sipe2000,MorimotoNagaosa2016Science,AhnNagaosaVishwanath2022}.
One route to enhancing SHG signals is to consider low-dimensional systems, where stronger light-matter coupling can lead to amplified resonant signals compared with bulk crystals.
%As a second-order response, SHG is highly sensitive to crystal symmetry and electronic structure and is widely used to probe broken inversion symmetry and crystalline defects.
%Furthermore, geometric quantities such as Berry connection and quantum metric can play an important role in the resonant SHG signal~\cite{Sipe2000,MorimotoNagaosa2016Science,AhnNagaosaVishwanath2022}.
%In low-dimensional materials, reduced dielectric screening and enhanced light-matter coupling can amplify SHG responses compared with their bulk counterparts.

Two-dimensional (2D) materials therefore provide a natural platform for observing strong SHG.
For example, transition metal dichalcogenides exhibit pronounced SHG signals arising from their intrinsic lack of inversion symmetry~\cite{KhanRahmanLu2022,HuangChenZhai2024,KumarLouZhao2013,SeylerYaoXu2015,WangWangZhang2019,BhallaCulcerAgarwal2022,BiswasYuanLow2023,LiangLiuYu2017,MennelPaurMueller2019,BeachLuckingTerrones2020,LiHeinz2013,HsuChouChang2014,YaoBasovSchuck2022,TangWangCao2026,CarvalhoMalard2019,TrolleSiefertPedersen2014,WangBalocchiUrbaszek2015,ShreeUrbaszekParadisanos2021,KimTaniguchiRyu2023}.
Among 2D systems, moir\'{e} materials provide a versatile platform for studying the SHG process, which will be the focus of this paper.

Moir\'{e} materials are formed by stacking and twisting two sheets of 2D materials, producing a long-wavelength interference pattern known as a moir\'{e} superlattice.
This superlattice can strongly modify the band structure, often giving rise to moir\'{e} flat bands with reduced bandwidth and an enhanced density of states (DoS)~\cite{BistritzerMacDonald2011,CaoJarilloHerrero2018}.
The large DoS allows multiple optical transitions to occur within a narrow frequency window, leading to strong resonant SHG signals.

SHG in moir\'{e} materials, where inversion symmetry is naturally broken by the interface structure, has been explored in both theory~\cite{LiuDai2020,ZhangLuLiu2022,YangHeZhang2024,RosasMendozaBarraza2024,YangYangFeng2025} and experiment~\cite{HsuChouChang2014,SongTongZhang2024,PsilodimitrakopoulosKioseoglouStratakis2019,YangYaoYang2020,ParadisanosMarieUrbaszek2022,StepanovNovolesovKatsnelson2020}.
In particular, enhanced SHG signals at selected twist angles have been reported and commonly attributed to the presence of moir\'{e} flat bands.
Despite these advances, the microscopic origin of the enhancement and its dependence on experimentally tuneable parameters remain only partially understood.
Previous studies have largely focused on limited parameter regimes, and a systematic characterisation of the SHG response as a function of twist angle, Fermi energy, and vertical bias voltage has not yet been fully explored.
In addition, the role of multi-step optical processes, such as double-resonance effects, has not been clarified in moir\'{e} systems.
In this work, we address these issues by presenting a comprehensive analysis of the SHG response in moir\'{e} systems.
We resolve the magnitude, phase, and resonance structure of the nonlinear susceptibility across a wide parameter space, and demonstrate that the SHG spectrum is governed not only by flat-band enhancement but also by pronounced double-resonance processes that can be efficiently tuned by external control parameters.

%SHG in moir\'{e} materials, which naturally break inversion symmetry at the interface, has been explored in both theory~\cite{LiuDai2020,ZhangLuLiu2022,YangHeZhang2024,RosasMendozaBarraza2024,YangYangFeng2025} and experiment~\cite{HsuChouChang2014,SongTongZhang2024,PsilodimitrakopoulosKioseoglouStratakis2019,YangYaoYang2020,ParadisanosMarieUrbaszek2022,StepanovNovolesovKatsnelson2020}, where enhanced signals at particular twist angles have been reported.
%These studies highlight the importance of the moir\'{e} flat bands for strong SHG resonance.
%However, the dependence of the SHG spectrum on experimentally tuneable parameters such as twist angle, Fermi energy, and vertical bias voltage has not yet been systematically examined.
%In this work, we present such a study, analysing the magnitude, phase, and resonance structure of the SHG response as a function of these control parameters.

\begin{figure}[t]
    \centering
    \includegraphics[width=8.5cm]{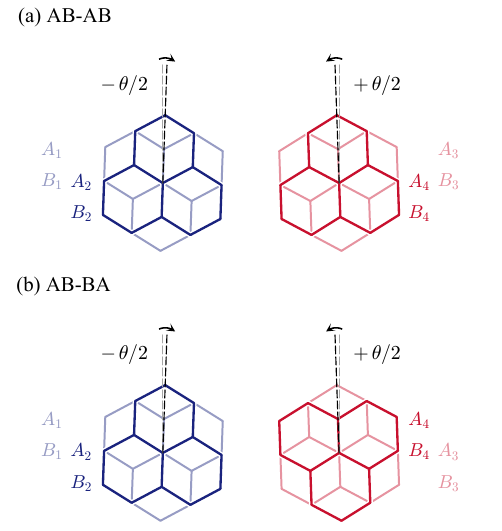}
    \caption{
    Stacking configurations of (a) AB--AB-stacked and (b) AB--BA-stacked twisted double bilayer graphene (TDBG).The lower BLG (blue) and the upper BLG (red) are rotated by $\mp\theta/2$ and stacked on top of each other.
}
\label{fig:1-geometry}
\end{figure}

As a concrete example, we consider twisted double bilayer graphene (TDBG), formed by stacking two AB-stacked bilayer graphene sheets with a relative twist, as illustrated 
in Fig.~\ref{fig:1-geometry}
~\cite{LiuDai2020,LiuDai2019,Koshino2019,ChebroluJung2019,BurgMacDonald2019,CrosseMoon2020,LiuKim2020,HaddadiOleg2020,CulchacMorell2020,ShenZhang2020,CaoJarilloHerrero2020,HeYankowitz2021,SzentpeteriMakk2021,WangTuctuc2022,Ma2022,RubioVerduPasupathy2022,TomicEnsslin2022,JoyaKawakamiKoshino2025}.
Owing to the absence of in-plane twofold rotational symmetry $C_{2z}$, TDBG supports a finite in-plane SHG response. In addition, a perpendicular electric displacement field enables continuous tuning of band gaps, making TDBG a versatile platform for controlling resonant optical processes. Furthermore, TDBG admits two stacking configurations, AB--AB and AB--BA, distinguished by a relative $180^\circ$ rotation between the two bilayers. Although these variants exhibit similar band structures, they possess distinct symmetries and topological properties~\cite{Koshino2019,ChebroluJung2019,LiuDai2019}, providing a natural setting to isolate stacking-dependent SHG responses with minimal influence from detailed band-structure variations.

Here we present a systematic investigation of the SHG spectrum in TDBG as a function of twist angle, Fermi energy, vertical bias, and stacking configuration. We find that the SHG response is strongly enhanced with decreasing twist angle, reflecting the emergence of moir\'{e} flat bands and the associated increase in the joint density of states. 
At small twist angles, the SHG signal is enhanced by one to two orders of magnitude compared with non-moir\'{e} AB-stacked bilayer graphene, which we compute as a reference system.

We further show that moir\'{e} flat bands qualitatively modify the resonance structure of the SHG response.
In particular, the reduced bandwidth enables a strong enhancement of the SHG response through a double-resonant mechanism, in which optical transitions at both $\omega$ and $2\omega$ are simultaneously satisfied over extended regions of the moir\'{e} Brillouin zone, resembling the enhancement found in confined systems with discrete energy levels~\cite{RosencherBoisNagle1996,NevouGuilllotMonroy2006,WuFangChen2014,FrigerioVirgilioOrtolani2021}.
By contrast, in conventional dispersive bands, the resonance condition is typically satisfied only in limited regions of momentum space, as we confirm for non-moir\'{e} AB-stacked bilayer graphene. 
Lastly, we demonstrate that the AB--AB and AB--BA configurations exhibit a systematic relative $\pi$ phase shift in the SHG response, providing a clear signature of their distinct symmetry and electronic structure.

The remainder of this paper is organised as follows.
In Sec.~\ref{sec:model}, we introduce the effective continuum Hamiltonians for AB-stacked bilayer graphene and TDBG and outline the theoretical formalism used to compute the SHG response.
In Sec.~\ref{sec:bilayer}, we analyse the SHG signal of AB-stacked bilayer graphene as a reference.
Sec.~\ref{sec:results} presents our results for TDBG, focusing on the roles of twist angle, Fermi energy, vertical bias, and stacking configuration.
Finally, we summarise our findings in Sec.~\ref{sec:conclusion}.

%%%%%%%%%%%%%%%%%%%%%%%%%%%%%%%%%%%%%%%%%%%%%%%%%%%%%%%%%%%%%%%%%%%%%%%%%%%%%%%%%%%%%%%%%%%%%%%%%%%%%%%%%%%%%%%%%%%%%%%
\section{Model and Methods}\label{sec:model}
\begin{figure}[t]
    \centering
    \includegraphics[width=8.5cm]{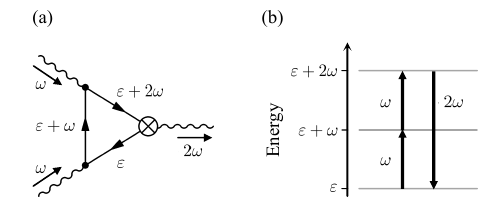}
    \caption{(a) Feynman diagram that is relevant to AB-stacked bilayer graphene and TDBG. The dots represent the interacting Hamiltonians and the crossed circle represents the current density operator. (b) Microscopic three-step optical transition that generates SHG.}
    \label{fig:1-feynman}
\end{figure}

In this section, we introduce the continuum model for AB-stacked bilayer graphene and its extension to twisted double bilayer graphene (TDBG), followed by the formalism used to compute the second-harmonic generation (SHG) response.

\subsection{AB-stacked bilayer graphene}\label{subsec:blg}
AB-stacked bilayer graphene consists of two graphene sheets with a relative in-plane shift such that an A site in one layer lies directly above a B site in the other.
%These vertically aligned sites form so-called dimerised pairs and are the origin of the in-plane polarisation of the system.
The primitive lattice vectors are $\vec{a}_1=a(1,0)$ and $\vec{a}_2=a(1/2,\sqrt{3}/2)$ with lattice constant $a=$~0.246~nm.
The corresponding reciprocal lattice vectors $\vec{b}_1=4\pi/\sqrt{3}a(\sqrt{3}/2,-1/2)$ and $\vec{b}_2=4\pi/\sqrt{3}a(0,1)$ and the corresponding Brillouin zone.
The Brillouin zone corners can be classified in to two valleys, $K_\pm$, around which we can expand a tight-binding Hamiltonian to obtain an effective Hamiltonian per valley.
The effective Hamiltonian $H_{\rm AB}$ is given by the following $4\times4$ matrix
\begin{equation}
        H_{\rm{AB}} = 
    \begin{pmatrix}
        H(\vec{k}) & g^\dagger(\vec{k})
        \\
        g(\vec{k}) & H^\prime(\vec{k})
    \end{pmatrix}
    +V_{\rm{AB}},
    \label{eq:abhamiltonian}
\end{equation}
where the corresponding $2\times2$ block matrices are defined as
\begin{align}
    \begin{split}
        H(\vec{k}) &=
        \begin{pmatrix}
            0 & -\hbar v_0k_-
            \\
            -\hbar v_0k_+ & \Delta^\prime
        \end{pmatrix},
        \\
        H^\prime(\vec{k}) &=
        \begin{pmatrix}
            \Delta^\prime & -\hbar v_0k_-
            \\
            -\hbar v_0k_+ & 0
        \end{pmatrix},
        \\
        g(\vec{k}) &=
        \begin{pmatrix}
            \hbar v_4k_+ & \gamma_1
            \\
            \hbar v_3k_- & \hbar v_4k_+
        \end{pmatrix}.
    \end{split}
\end{align}
The velocity parameters $v_i$ are defined from the hopping parameters $\gamma_i$ as $v_i= \sqrt{3}a\abs{\gamma_i}/(2\hbar)$, $\gamma_i$, and $k_{\pm}=\xi k_x\pm ik_y$ with the valley index $K_\xi$.
Then, $\gamma_0=-2.4657$~eV is the intralayer nearest-neighbour hopping, $\gamma_1=$~0.4~eV is the hopping between the dimerised sites, $\gamma_3=$~0.32~eV and $\gamma_4=$~0.044~eV are the interlayer diagonal hopping, and $\Delta^\prime=$~0.050~eV is the on-site potential at the dimerised sites.
Finally, the $V_{\rm AB}$ term models the vertical bias that can be applied across the two layers, and can be written as
\begin{equation}
       V_{\rm{AB}} = 
    \begin{pmatrix}
        \frac{1}{2}\Delta\,\mathbb{I} & 
        \\
        & -\frac{1}{2}\Delta\,\mathbb{I}
    \end{pmatrix}.
    \label{eq:blgv}
\end{equation}

\subsection{Twisted double bilayer graphene}\label{subsec:hamiltonian}
Twisted double bilayer graphene (TDBG) is constructed by stacking two sheets of AB-stacked bilayer graphene with a relative twist.
The stacking can happen in two ways: with the two bilayers aligned in the same direction, or by introducing a $180^\circ$ offset to the second layer.
The configuration given by the first method is known as the AB--AB stack, while the other is known as the AB--BA stack.
The two TDBG variants are known to have very similar band structures despite the difference in the stacking order and symmetry~\cite{Koshino2019,ChebroluJung2019,LiuDai2019}.
%Nevertheless, the different symmetry property leads to completely different valley Chern numbers.

Then, the relative twist is performed by rotating the lower ($l=1$) and upper ($l=2$) bilayers through an angle of $\mp\theta/2$.
Through this rotation, the reciprocal lattice vectors of the individual bilayers are also rotated as  $\vec{b}^{(l)}_i = R\left[(-1)^l\theta/2\right]\vec{b}_i$, where $R(\theta)$ is the rotation matrix.
As a consequence, Brillouin zone corners $\vec{K}_\xi$ are also rotated to $\vec{K}^{(l)}_\xi$.
We can take advantage of this to define a zone-folded moir\'{e} Brillouin zone, with moir\'{e} reciprocal lattice vectors $\vec{G}^{\rm{M}}_i=\vec{b}^{(1)}_i-\vec{b}^{(2)}_i$, where the zone corners are given by the $\vec{K}^{(l)}_\xi$.
Especially, at small twist angles $\theta$, the system shows a large-scale moir\'{e} pattern with a lattice constant $L^{\rm M} \sim a/\theta$, which allows us to coarse-grain the system to the moir\'{e} scale and ignore the underlying atomic lattice, leading us to define an effective continuum Hamiltonian.

To construct the effective continuum Hamiltonian for TDBG, we can build on the effective Hamiltonian for bilayer graphene \cite{Koshino2019}.
By introducing the interlayer moir\'{e} interaction
\begin{align}
    U(\vec{r}) = 
    \begin{pmatrix}
        u & u^\prime
        \\
        u^\prime & u
    \end{pmatrix}
    & +
    \begin{pmatrix}
        u & u^\prime\omega^{-\xi}
        \\
        u^\prime\omega^\xi & u
    \end{pmatrix}
    e^{i\xi\vec{G}^{\rm{M}}_1\vdot\vec{r}}
    \\
    \nonumber
    & +
    \begin{pmatrix}
        u & u^\prime\omega^\xi
        \\
        u^\prime\omega^{-\xi} & u
    \end{pmatrix}
    e^{i\xi\left(\vec{G}^{\rm{M}}_1+\vec{G}^{\rm{M}}_2\right)\vdot\vec{r}},
\end{align}
the total AB--AB and AB--BA Hamiltonians at valley $\xi$ can be given as
\begin{align}
    \begin{split}
        H_{\rm{AB\text{--}AB}} &= 
        \begin{pmatrix}
            H(\vec{k}_1) & g^\dagger(\vec{k}_1) & &
            \\
            g(\vec{k}_1) & H^\prime(\vec{k}_1) & U^\dagger(\vec{r}) &
            \\
            & U(\vec{r}) & H(\vec{k}_2) & g^\dagger(\vec{k}_2)
            \\
            & & g(\vec{k}_2) & H^\prime(\vec{k}_2)
        \end{pmatrix} + V,
        \\
        H_{\rm{AB\text{--}BA}} &= 
        \begin{pmatrix}
            H(\vec{k}_1) & g^\dagger(\vec{k}_1) & &
            \\
            g(\vec{k}_1) & H^\prime(\vec{k}_1) & U^\dagger(\vec{r}) &
            \\
            & U(\vec{r}) & H^\prime(\vec{k}_2) & g(\vec{k}_2)
            \\
            & & g^\dagger(\vec{k}_2) & H(\vec{k}_2)
        \end{pmatrix} + V.
    \end{split}
    \label{eq:tdbghamiltonian}
\end{align}
Here,  $\vec{k}_l=R\left[(-1)^l\theta/2\right](\vec{k}-\vec{K}^{(l)}_\xi)$, and the vertical bias is represented by the matrix
\begin{equation}
    V = 
    \begin{pmatrix}
        \frac{3}{2}\Delta\,\mathbb{I} & & &
        \\
        & \frac{1}{2}\Delta\,\mathbb{I} & &
        \\
        & & -\frac{1}{2}\Delta\,\mathbb{I} &
        \\
        & & & -\frac{3}{2}\Delta\,\mathbb{I}
    \end{pmatrix}.
\end{equation}
If we pay attention to the structure of the two Hamiltonians, we can see that the diagonal blocks are the bilayer Hamiltonians, with the exception of the lower-right block in $H_{\rm{AB\text{--}BA}}$, which reflects the extra $180^\circ$ rotation of the bilayer.
When numerically evaluating these Hamiltonians, we introduce a cutoff on the Fourier expansions of the moir\'{e} interaction $U(\vec{r})$.
In the present work, we have adopted a cutoff of 
%$\norm{\vec{k}}\leq 5\,\norm{\vec{G}^{\rm{M}}_i}$ within $\vec{k}$-space.
$|\vec{k}|\leq 5\,|\vec{G}^{\rm{M}}_i|$ within $\vec{k}$-space.

Finally, we discuss the symmetries of the two Hamiltonians. In both variants, the system shows the in-plane threefold rotational symmetry $C_{3z}$ inherited from the original graphene.
In addition, in the absence of a vertical bias, the AB--AB TDBG has an out-of-plane twofold rotational symmetry along the $x$-axis $C_{2x}$, while the AB--BA type has the same symmetry along the $y$-axis $C_{2y}$.

\subsection{Second-harmonic generation}\label{subsec:shg}
Second-harmonic generation (SHG) is a second-order optical response, in which light with frequency $\omega$ is transformed into a photocurrent with frequency $2\omega$; a phenomenon known as frequency doubling.
The photocurrent $j_\mu$ in the $\mu$ direction is expressed as
\begin{equation}
    j_\mu(2\omega) = \sigma^\mu_{\alpha\alpha}(2\omega;\omega,\omega)\;E_\alpha(\omega)E_\alpha(\omega),
    \label{eq:photocurrent}
\end{equation}
where $\sigma^\mu_{\alpha\alpha}$ is the SHG conductivity tensor and $E_\alpha$ is the electric field of the beam of light polarised in the $\alpha$ direction.
In the present geometry, a linearly polarised beam is incident perpendicularly to the plane of the sheet, and we therefore adopt $\mu,\alpha=x,y$.
Importantly, a finite SHG signal requires the absence of the inversion symmetry of the crystal.
This can be checked by inspecting Eq.~\eqref{eq:photocurrent}, where it can be seen that the parities under inversion of either sides of the equation are opposite, thus, leading to a vanishing $\sigma^\mu_{\alpha\alpha}$ in centrosymmetric systems.
In addition, for 2D systems, the in-plane twofold rotational symmetry $C_{2z}$ acts against SHG to cancel the signal.

To evaluate the $\sigma^\mu_{\alpha\alpha}$ tensor, perturbative methods can be employed, where the electric field is introduced to the Hamiltonian through the minimal coupling substitution~\cite{ParkerMorimoto2019}.
This method corresponds to the velocity gauge, in which the nonlinear optical conductivity tensors take numerically convenient forms, particularly when the velocity matrix elements are known.
%This method corresponds to using a gauge known as the velocity gauge, in which the nonlinear optical conductivity tensors can be written in numerically convenient forms, especially when the velocity matrix are known.
Furthermore, taking advantage of the $\vec{k}$-linear form of the Hamiltonian Eq.~\eqref{eq:tdbghamiltonian} under interest, we can neglect terms including second- and third-order derivatives of the Hamiltonian.
Then, the contribution to the SHG conductivity tensor is represented by the Feynman diagram shown in Fig.~\ref{fig:1-feynman}~(a), which reads~\cite{ParkerMorimoto2019}
\begin{align}
    &\sigma^\mu_{\alpha\alpha}(\omega) = \frac{e^3}{2\pi^2\omega^2}\int d^2k\sum_{a,b,c}\frac{v^\mu_{ac}v^\alpha_{cb}v^\alpha_{ba}}{\varepsilon_{ba}-\varepsilon_{cb}}\;\times \nonumber
    \\
    &\left(\frac{f_{ab}}{\hbar\omega-\varepsilon_{ba}+i\eta}+\frac{f_{bc}}{\hbar\omega-\varepsilon_{cb}+i\eta}-\frac{2f_{ac}}{2\hbar\omega-\varepsilon_{ca}+2i\eta}\right).
    \label{eq:shgconducutivity}
\end{align}
Here, $e (>0)$ is the magnitude of the electron charge, $\varepsilon_{ba} = \varepsilon_{b} - \varepsilon_{a}$, where $\varepsilon_{a}$ is the eigenenergy of the Bloch state $\ket{a}$, $f_{ab} = f(\varepsilon_a) - f(\varepsilon_b)$, where $f(\varepsilon)$ is the Fermi-Dirac occupation function, $v^\alpha_{ba}$ are the matrix elements of the velocity operators $ v^\alpha = \hbar^{-1}\partial H/\partial k_\alpha$.
Finally, $\eta$ is the phenomenological broadening parameter that is introduced via the analytical continuation of the Matsubara frequencies.
To be explicit, we replace the Matsubara frequencies $i\omega$ and $2i\omega$  to real frequencies $\hbar\omega+i\eta$ and $2\hbar\omega+2i\eta$, respectively.
The the factor of $2$ in $2i\eta$ assumes that the phenomenological broadening originates from the adiabatic switch on of the external field.
We justify this in Appendix~\ref{appx:comparison}, where we compare the numerical results from Eq.~\ref{eq:shgconducutivity} and an alternative expression using the electron broadening and show that the results coincide numerically.

Finally, we reveal the finite and independent components of the $\sigma^\mu_{\alpha\alpha}$ tensor by imposing the crystal symmetry of TDBG.
As we have discussed previously, both variants of TDBG host the $C_{3z}$ symmetry, which restricts the SHG tensor $\sigma^\mu_{\alpha\alpha}$ to two independent components, $\sigma^x_{xx}$ and $\sigma^y_{yy}$.
The other components are related via~\cite{BrunPedersen2015,LiuDai2020}
\begin{align}
    \begin{split}
        \sigma^x_{xx} &= -\sigma^y_{xy} = -\sigma^y_{yx} = -\sigma^x_{yy},
        \\
        \sigma^y_{yy} &= -\sigma^x_{xy} = -\sigma^x_{yx} = -\sigma^y_{xx}.
    \end{split}
    \label{eq:symmetry}
\end{align}
When the vertical bias is turned off, the out-of-plane rotational symmetry is recovered, which further restricts the $\sigma^\mu_{\alpha\alpha}$.
For AB--AB, we find that the $C_{2x}$ symmetry renders $\sigma^y_{yy} = 0$, and the only remaining component is $\sigma^x_{xx}$.
Whereas, in AB--BA, the $C_{2y}$ symmetry forces $\sigma^x_{xx} = 0$ and $\sigma^y_{yy}$ to be finite.

\subsection{Double resonance}\label{subsec:shg_schematic}
\begin{figure}[t]
    \centering
    \includegraphics[width=8.5cm]{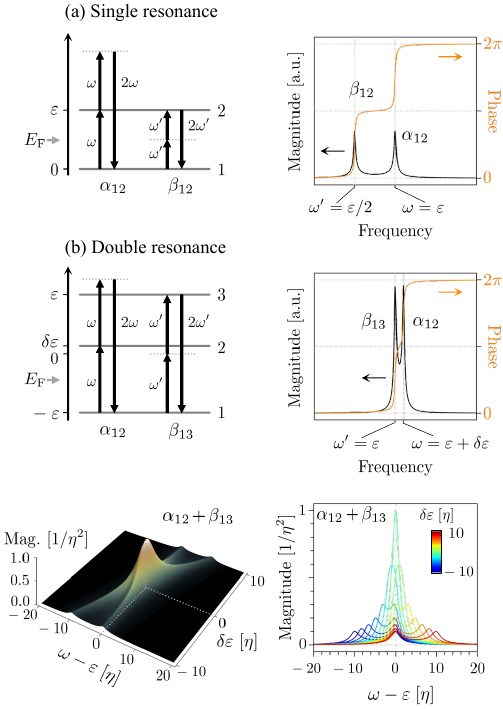}
    \caption{Schematic diagrams of the processes involved in the SHG signal. (a) Single resonances via one-photon ($\alpha_{12}$) and two-photon ($\beta_{12}$) processes between two energy levels, shown on the left. The $\alpha_{12}$ and $\beta_{12}$ processes produce a peak at $\omega=\varepsilon$ and $\omega^\prime=\varepsilon/2$ respectively. (b) Top two panels show the situation approaching double resonance involving three levels, where the $\alpha_{12}$ and $\beta_{13}$ processes produce a peak at $\omega=\varepsilon$ as $\delta\varepsilon\to0$. Bottom two panels are the plots of the double-resonant signal as a function of $\delta\varepsilon$. It can be seen that the signal is significantly enhanced as $\delta\varepsilon$ approaches 0. The family of curves is universal, expressed in units involving the broadening parameter $\eta$.}
    \label{fig:2-shg_schematic}
\end{figure}

In this section, we examine the resonance structure of the SHG response through a series of schematic examples.
%In this section, we will take our time to understand the resonance structure of SHG signals by studying its behaviour in schematic examples.
The SHG is composed of three optical process: two excitations via energy $\omega$ and a relaxation via energy $2\omega$, as shown in Fig.~\ref{fig:1-feynman}(b).
%We refer to the respective processes as 1-photon ($\alpha$) and 2-photon ($\beta$) processes, and we achieve resonance once we match two of the three levels with real energy bands.

We begin by the simplest case, where we consider two levels separated by energy $\varepsilon$, represented by the solid thick lines on the left panel of Fig.~\ref{fig:2-shg_schematic}(a).
As the SHG process involves three states, illustrated in Fig.~\ref{fig:1-feynman}(b), the process will always involve the two real levels (thick solid lines) and one off-resonant virtual level (thin line).
Then, in this situation, we have two possible single resonances, labelled $\alpha_{12}$ and $\beta_{12}$.
The $\alpha_{12}$ process involves a real excitation between the the two real levels via energy $\omega$, thus, giving rise to a peak at $\omega=\varepsilon$, as shown on the right panel.
As such, we refer to such transitions as 1-photon processes.
On the other hand, the $\beta_{12}$ process is resonant via the $2\omega^\prime$ transition, which leads to a peak at $\omega^\prime=\varepsilon/2$, i.e., at half the size of the energy gap.
Since the process involves two successive excitations, aided by a virtual level, we call this a 2-photon process.
Herein, we will label 1- and 2-photon resonances from level $i$ to level $j$ as $\alpha_{ij}$ and $\beta_{ij}$, respectively.
We also plot the phase of the signal in orange, where each peak picks up a phase of $\pi$.

Let us now consider the situation where we have three levels that are almost equidistant in energy, as shown in the top left panel of Figure~\ref{fig:2-shg_schematic}~(b).
Here, the levels~1, 2, and 3 are set to energies $-\varepsilon$, $\delta\varepsilon$, and $\varepsilon$, respectively, with the Fermi energy placed between levels~1 and 2.
In this case, we are particularly interested in the one-photon process $\alpha_{12}$ and the two-photon process $\beta_{13}$; the former produces a peak at $\omega=\varepsilon+\delta\varepsilon$, while the latter at $\omega^\prime=\varepsilon$, as shown in the top right panel.
As $\delta\varepsilon \to 0$, the $\alpha_{12}$ and $\beta_{13}$ resonances approach one another and eventually coalesce, producing a single, strongly enhanced feature in the SHG spectrum, as shown in the bottom two panels.
This situation, in which the one-photon and two-photon channels are simultaneously activated, is referred to as double resonance.
Importantly, the resulting peak is not simply the sum of two independent Lorentzian peaks; instead, the simultaneous activation of the two channels leads to a qualitatively different line shape.
The enhancement stems from the energy denominator $\varepsilon_{ba}-\varepsilon_{cb}$ appearing in Eq.~\eqref{eq:shgconducutivity}, which vanished when the three levels are exactly equidistant.

We can describe the limiting behaviour of the SHG signal in the vicinity of the double resonance.
Noting that the entire signal will be dominated by the triple of states $(a,b,c)$ causing the double resonance, for small $\delta\varepsilon$, $\sigma^\mu_{\alpha\alpha}$ is given as
\begin{align}
    &\sigma^\mu_{\alpha\alpha} \sim \frac{e^3}{4\pi^2\omega^2}\int d^2k\,\left(f_{ab} - f_{bc}\right)\frac{v^\mu_{ac}v^\alpha_{cb}v^\alpha_{ba}}{(\hbar\omega-\varepsilon+i\eta)^2}+\mathcal{O}(\delta\varepsilon).
    \label{eq:shg_dr}
\end{align}
We can see that the leading order is in fact independent of $\delta\varepsilon$ and the apparent divergence by the $(\varepsilon_{ba}-\varepsilon_{cb})^{-1}$ factor is explicitly removed.
Moreover, it is clear that the line-shape is now the square of a Lorentzian, which explains the origin of the strong enhancement~\cite{RosencherBoisNagle1996,FrigerioVirgilioOrtolani2021}.
The details of the derivation is presented in Appendix~\ref{appx:dr_analysis}.

%%%%%%%%%%%%%%%%%%%%%%%%%%%%%%%%%%%%%%%%%%%%%%%%%%%%%%%%%%%%%%%%%%%%%%%%%%%%%%%%%%%%%%%%%%%%%%%%%%%%%%%%%%%%%%%%%%%%%%%
\begin{figure*}[ht]
    \centering
    \includegraphics[width=15cm]{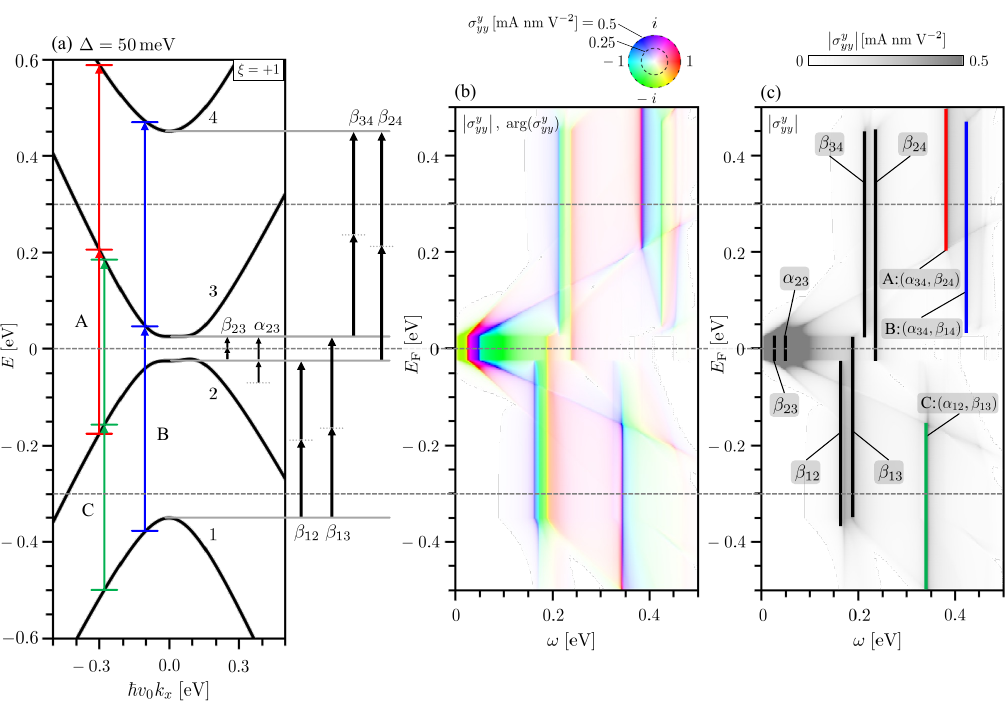}
    \caption{(a) Band structure of non-moir\'{e} AB-stacked bilayer graphene from the $K_+$ valley. A vertical bias voltage of $\Delta=50$~meV is applied across the two layers. (b) Magnitude and phase of the SHG tensor, $\abs{\sigma^y_{yy}}$ and $\arg(\sigma^y_{yy})$, as a function of $(\omega;E_{\rm F})$ on a colour-density plot, where we plot $\omega$ and $E_{\rm F}$ on the horizontal and vertical axes. Namely, $\arg(\sigma^y_{yy})$ is given by the hue while $\abs{\sigma^y_{yy}}$ is given by the saturation. We note that the scale of the vertical axes of the band structure and the density maps are aligned. (c) Colour-density plot of the magnitude $\abs{\sigma^y_{yy}}$ with indications of the important spectral features.}
    \label{fig:3-blg}
\end{figure*}

\section{SHG in AB-stacked bilayer graphene}\label{sec:bilayer}

\begin{figure}[t]
    \centering
    \includegraphics[width=8cm]{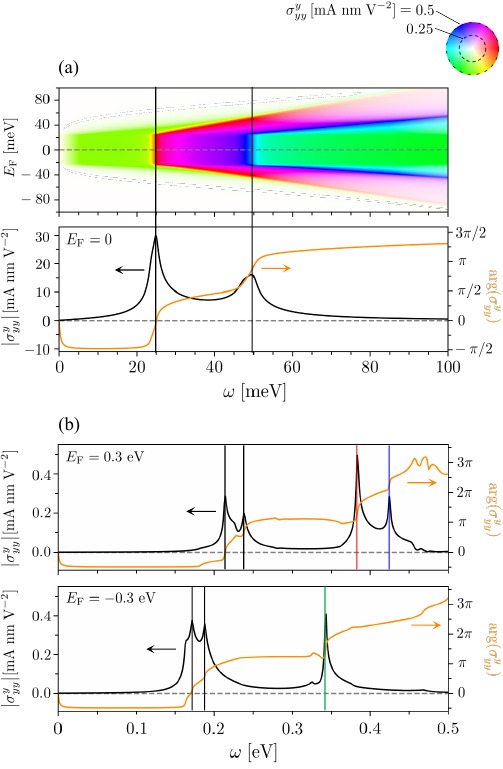}
    \caption{(a) Zoomed-in view of the colour-density plot near the CNP shown in the top panel. The bottom panel shows the slice through $E_{\rm F}=0$, the where magnitude and phase are plotted in black and orange, respectively. The vertical lines are the $\alpha_{23}$ and $\beta_{23}$ peaks indicated in Fig.~\ref{fig:3-blg}~(c). (b) Top (bottom) panel is a slice through $E_{\rm F}=0.3$~eV ($E_{\rm F}=-0.3$~eV), where, again, the vertical lines correspond to the features indicated in Fig.~\ref{fig:3-blg}~(c).}
    \label{fig:3-blg_zoom}
\end{figure}

In this section, we give an overview of the second-harmonic generation (SHG) in non-moir\'{e} AB-stacked bilayer graphene as a pre-requisite to understand the results in the moir\'{e} system.
The results are shown in Figs.~\ref{fig:3-blg} and \ref{fig:3-blg_zoom}.
In Fig.~\ref{fig:3-blg}(a), we show the band structure of bilayer graphene from the $K_+$ valley with a vertical bias voltage $\Delta=50$~meV, which opens a gap of 50~meV at the charge-neutral point (CNP).
The bands are labelled from 1 to 4 in ascending order of energy.
The bias voltage is applied to break the inversion symmetry of the system, allowing it to be SHG-active.
Due to the mirror symmetry $M_x$ across the $y\text{--}z$ plane, the only surviving component of the tensor is $\sigma^y_{yy}$, which, in general, would be a complex quantity calculated using Eq.~\eqref{eq:shgconducutivity}.

Fig.~\ref{fig:3-blg}(b) shows the SHG conductivity tensor $\sigma^y_{yy}(\omega;E_{\rm F})$ as a function of the light frequency $\omega$ and Fermi energy $E_{\rm F}$ on a colour-density map.
The horizontal axis corresponds to $\omega$, while the vertical axis corresponds to $E_{\rm F}$ which is aligned with the energy scale of the band-structure plot.
Each horizontal slice therefore represents a $\sigma^y_{yy}$--$\omega$ spectrum at fixed $E_{\rm F}$.  
The complex response is visualised by mapping the magnitude $|\sigma^y_{yy}|$ to the saturation and the phase $\arg(\sigma^y_{yy})$ to the hue of each pixel, according to the colour wheel above the figure.
Finally, Fig.~\ref{fig:3-blg}(c) shows the same data with the phase information removed, showing only $|\sigma^y_{yy}|$, with the main spectral features annotated for clarity.

Inspecting the colour-density plots, we find a strong response near the CNP.
These are the one-photon $\alpha_{23}$ and two-photon $\beta_{23}$ resonances between the band endge of bands 2 and 3, occurring at $\omega\approx50$~meV and 25~meV, respectively.
The sharpness of the peaks originates from the flatness of the bands around the $K_{\pm}$ points, where a large joint density of states (JDoS) enhances the response.
The nature of the double peak can be seen more clearly in Fig.~\ref{fig:3-blg_zoom}(a), where we show a zoomed-in image of the phase plot near the CNP together with a slice of the SHG strength (black) and phase (orange) along the $E_{\rm F}=0$ line.
%On the density plot of Fig.~\ref{fig:4-blgshg_zoom}~(a), we can see that the phase suddenly jumps at $\omega=25$~meV and 50~meV, which can be clearly seen in the bottom panel.
%This phase behaviour contrasts with linear response, where a single gap produces a phase change of $\pi$, whereas in the present SHG case the accumulated phase is $2\pi$.

Moving away from the CNP, two additional peaks appear above and below charge neutrality around $\omega\approx0.2$~eV.
%, which can be inferred from the phase jumps.
These are the two-photon processes involving the outermost bands, i.e., $\beta_{12}$, $\beta_{13}$, $\beta_{24}$, and $\beta_{34}$, indicated in the band structure.
The microscopic origin of these $\beta$ peaks can be better understood by inspecting the constant-$E_{\rm F}$ slices of the density plots in Figs.~\ref{fig:3-blg_zoom}(b) and (c).
At fixed $E_{\rm F}=0.3$~eV, shown in Fig.~\ref{fig:3-blg_zoom}(b), we find two features near 0.2~eV and 0.25~eV, corresponding to the $\beta_{34}$ and $\beta_{24}$ processes, respectively.
The sharp peaks, indicated by the black vertical lines in Fig.~\ref{fig:3-blg}(c), are due to the divergence of the JDoS between the band edges and are therefore activated when $E_{\rm F}$ lies between the band edges of bands 3 and 4.
A similar mechanism applies for $E_{\rm F}=-0.3$~eV, where the two peaks in Fig.~\ref{fig:3-blg_zoom}(c) are the signals from the $\beta_{13}$ and $\beta_{12}$ processes.

\begin{comment}
\blue{
Furthermore, moving to higher frequencies at $E_{\rm F}=0.3$~eV, we find features around $\omega\approx0.4$~eV.
In this frequency region, there is a coexistence of the 1-photon and 2-photon processes, where the signals from the two processes interfere in a complex manner.
Although the interference generally results in an uncharacteristic spectrum, the system generates a strong signal when the double-resonance condition is met.
Precisely, we find two prominent peaks, indicated by the red and blue vertical lines, which are signals from the double-resonant processes.
Resonance A arises from the simultaneous activation of the $\alpha_{34}$ and $\beta_{24}$ channels, which occurs when bands 2, 3, and 4 become equidistant at a specific $\vb*{k}$ point, as shown by the red arrows in Fig.~\ref{fig:3-blg}(a).
Similarly, resonance B is the double resonance between $\alpha_{34}$ and $\beta_{14}$, with bands 1, 3, and 4 being equidistant, indicated by the blue arrows.
While we also see double resonance for $E_{\rm F} < 0$ (resonance C) by a similar mechanism, we only find one resonance rather than two.
This is due to the broken particle-hole symmetry, and bands 1, 2, and 4 never meet the double-resonance condition.
}
\end{comment}

Furthermore, moving to higher frequencies at $E_{\rm F}=0.3$~eV, we find three prominent peaks, labelled A, B, and C, in the range $\omega \approx 350$--400~meV. These originate from double-resonant processes involving transitions among three approximately equidistant energy levels, as discussed in Sec.~\ref{subsec:shg}.
Resonance A arises from the simultaneous activation of the $\alpha_{34}$ and $\beta_{24}$ channels, which occurs when bands 2, 3, and 4 become equally spaced in energy at a given $\vb*{k}$ point. The condition $\varepsilon_{23}=\varepsilon_{34}$ is satisfied along a line in the Brillouin zone, with the common energy spacing varying along the line. Peak A occurs at a particular point where the constant-energy contours of $\varepsilon_{23}$ and $\varepsilon_{34}$ become tangent, thereby providing an enhanced phase space for the double resonance. This point is indicated by the red arrows in Fig.~\ref{fig:3-blg}(a).
Similarly, resonance B originates from the double resonance between the $\alpha_{34}$ and $\beta_{14}$ channels, for which bands 1, 3, and 4 become equally spaced in energy. The corresponding point is indicated by the blue arrows in Fig.~\ref{fig:3-blg}(a).
For $E_{\rm F}<0$, we also observe a double resonance (peak C) arising from an analogous mechanism. In this case, however, only one double-resonance appears, because the broken particle-hole symmetry prevents bands 1, 2, and 4 from satisfying the double-resonance condition.

\section{SHG in TDBG}\label{sec:results}

In this section, we present the numerical results of the second-harmonic generation (SHG) in AB--AB and AB--BA twisted double bilayer graphene (TDBG).
We begin by analysing the SHG response for a representative set of parameters (AB--BA stack, $\theta=0.8^\circ$, $\Delta=50$~meV) in order to establish the basic resonance structure of SHG in the presence of moir\'{e} flat bands.
We also compare the result with the response by non-moir\'{e} AB-stacked bilayer graphene from Sec.~\ref{sec:bilayer} to highlight the effects of the moir\'{e} interaction.
Having clarified the characteristic behaviour at a single parameter, we then study the intrinsic SHG response at zero vertical bias, $\Delta=0$, and examine its evolution as the twist angle $\theta$ is varied.
We subsequently investigate the effect of a finite vertical bias at fixed a twist angle, $\theta=0.8^\circ$, and finally compare the SHG responses between AB--AB and AB--BA stacking configurations.

Throughout this section, the SHG conductivity is computed using the numerical expression in Eq.~\eqref{eq:shgconducutivity} together with the continuum Hamiltonian in Eq.~\eqref{eq:tdbghamiltonian}.  
In computing this expression, we set a cutoff of 30 bands above and below the charge-neutral point (CNP), which we have verified to be sufficient for numerical convergence.

\begin{figure*}[t]
    \centering
    \includegraphics[width=17cm]{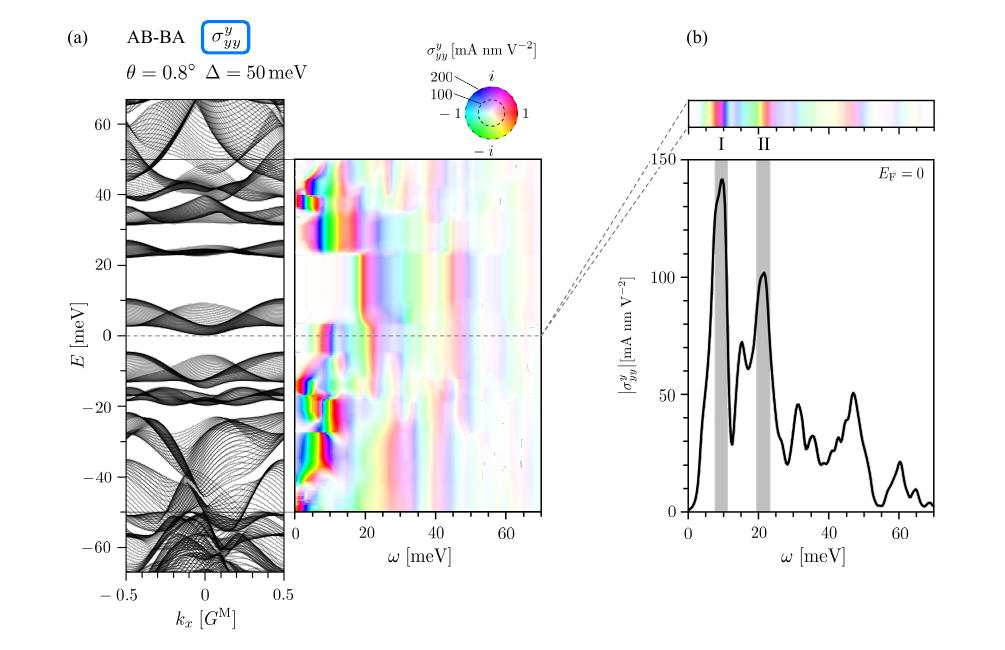}
    \caption{(a) Band structure of AB--BA stacked TDBG at $\theta=0.8^\circ$ and $\Delta=50$~meV, together with the corresponding SHG response $\sigma^y_{yy}(\omega;E_{\rm F})$ shown as a colour-density plot. The band structure is obtained by overlaying one-dimensional cuts at fixed $k_y$ over the moir\'{e} Brillouin zone. In the SHG density plot, the colour hue represents the phase of the complex conductivity tensor, while the brightness encodes its magnitude, as indicated by the colour wheel. The horizontal and vertical axes correspond to the excitation frequency $\omega$ and the Fermi energy $E_{\rm F}$, respectively, with the $E_{\rm F}$ axis aligned to the band structure. (b) Magnitude of $\sigma^y_{yy}(\omega;E_{\rm F}=0)$ as a function of frequency, with two dominant peaks labelled I and II.}
    \label{fig:4-theta04delta0}
\end{figure*}

\begin{figure}[t]
    \centering
    \includegraphics[width=8.5cm]{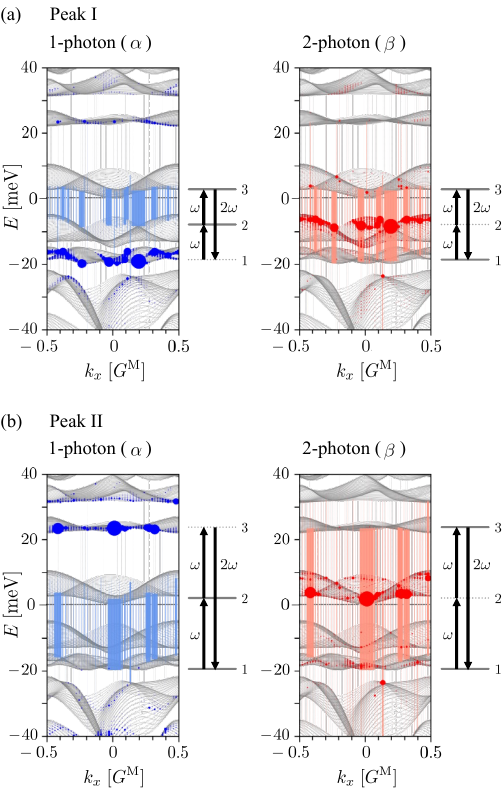}
    \caption{SHG contributions to (a) peak I and (b) peak II over $\vec{k}$-space. The one-photon ($\alpha$, blue) and two-photon ($\beta$, red) processes are shown separately. For each $(k_x,k_y)$ point, the triplet of states yielding the largest contribution is identified. The real states are connected by the vertical line while the virtual state is depicted by the filled circle. The thickness of the lines and the size of the circle indicate the relative weight of each contribution.}
    \label{fig:4-theta04delta0_dr}
\end{figure}

\subsection{Resonance structure}\label{subsec:resonant}

Figure~\ref{fig:4-theta04delta0}(a) shows the band structure and $\sigma^y_{yy}(\omega;E_{\rm F})$ response for AB--BA stacked TDBG at $\theta=0.8^\circ$ and $\Delta=50$~meV.
The band structure is obtained by superposing one-dimensional cuts at fixed $k_y$ over the moir\'{e} Brillouin zone, thereby visualising the full energy spectrum.
The SHG spectrum is presented as a colour-density plot in the $(\omega,E_{\rm F})$ plane, with the $E_{\rm F}$ axis aligned to the band structure for direct comparison.

To elucidate the microscopic origin of the SHG response, we first focus on the cut at $E_{\rm F}=0$ shown in Fig.~\ref{fig:4-theta04delta0}(b).
Two dominant peaks, labelled I and II, are clearly visible.
Compared with the corresponding spectrum of non-moir\'{e} AB-stacked bilayer graphene in Fig.~\ref{fig:3-blg}(a), shown over a similar frequency range, the peak intensity is enhanced by approximately a factor of five.
As we show below, this enhancement originates from a combination of moir\'{e}-band flattening and the emergence of abundant double-resonant optical processes.

To identify the processes responsible for peaks I and II, we perform a $\vb*{k}$-resolved decomposition of the SHG response.
For each $(k_x,\,k_y)$ point, we determine the triplet of states that yields the largest contribution within the shaded frequency windows.
The resulting distributions are shown in Fig.~\ref{fig:4-theta04delta0_dr}, where the one-photon
($\alpha$) and two-photon ($\beta$) channels are displayed separately.
In these plots, the initial and final states are connected by a line, while the virtual state is represented by a filled circle; both the line thickness and the circle size are scaled according to the relative contribution of the corresponding process.
The Fermi energy, set to $E_{\rm F}=0$, is indicated by a horizontal line.

For peak I [Fig.~\ref{fig:4-theta04delta0_dr}(a)], the dominant one-photon contribution arises from optical transitions between neighbouring bands across the Fermi level, while the corresponding virtual state lies in the band immediately below the initial state.
Labelling the relevant bands 1, 2, and 3 in order of increasing energy, this process corresponds to an $\alpha_{23}$ transition mediated by band~1,  as illustrated in the accompanying level diagram.
At the same frequency, the dominant two-photon contribution involves the same set of bands and corresponds to a $\beta_{13}$ transition with band~2 acting as the virtual state.
As a result, the one-photon and two-photon channels become simultaneously resonant, satisfying the double-resonance condition.
%The corresponding contributions are distributed over a broad region of the moir\'e Brillouin zone rather than being confined to a limited region in $\vb{k}$ space, indicating that the resonance contributes substantially to the observed SHG peak.
An analogous situation occurs for peak II [Fig.~\ref{fig:4-theta04delta0_dr}(b)], where the dominant contributions again arise from simultaneously resonant
$\alpha$ and $\beta$ processes.

Notably, unlike the case of AB-stacked bilayer graphene discussed in Sec.~\ref{sec:bilayer}, these double-resonant conditions are satisfied throughout extended portions of the moir\'e Brillouin zone rather than being confined to isolated regions of momentum space.
Consequently, such resonant processes in TDBG are neither rare nor fine tuned, but rather emerges as a generic feature of the moir\'e flat-band structure.
The origin of this behaviour can be traced to the formation of moir\'e flat bands.
Because neighbouring bands remain nearly parallel over wide regions of momentum space, the energy spacings required for the one-photon and two-photon resonances can be satisfied simultaneously for many $\vb*{k}$ points.
In addition, the large joint density of states associated with the flat bands further increases the number of available optical transitions within a narrow energy window, leading to a strong enhancement of the SHG response.
The resulting situation is analogous to nonlinear optical processes in systems with discrete energy levels, where double resonance is often exploited to enhance nonlinear optical responses~\cite{RosencherBoisNagle1996,NevouGuilllotMonroy2006,WuFangChen2014,FrigerioVirgilioOrtolani2021}.

The SHG response in TDBG therefore differs qualitatively from
that of conventional bilayer graphene.
Whereas resonant processes in bilayer graphene are confined to
limited regions of momentum space, the moir\'e flat bands of TDBG
support a large number of resonant channels distributed throughout
the Brillouin zone.
The resulting abundance of double-resonant processes provides a
general mechanism for the strong nonlinear optical responses
observed in flat-band moir\'{e} systems.

\subsection{Evolution with twist angle}\label{subsec:theta}
\begin{figure*}[t]
    \centering
    \includegraphics[width=17cm]{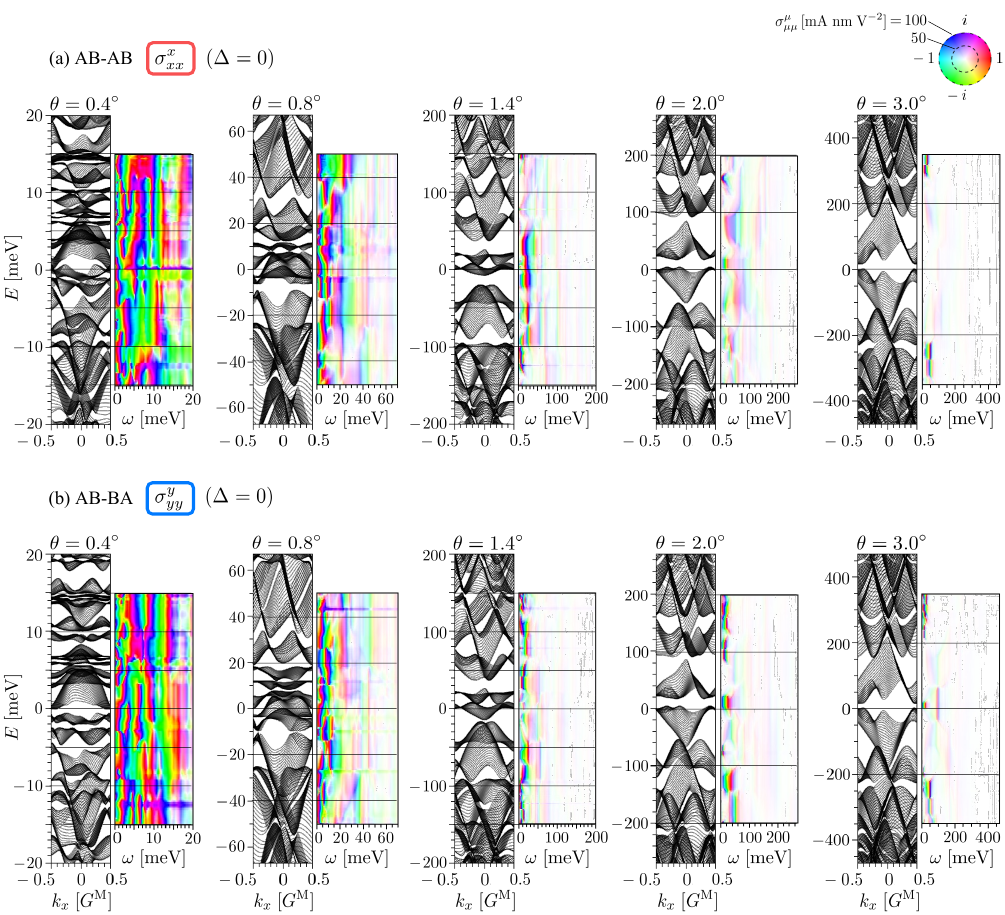}
    \caption{(a) Band structures of AB--AB stacked TDBG and colour-density plots of $\sigma^x_{xx}(\omega;E_{\rm F})$ for various twist angles $\theta=0.4^\circ$, $0.8^\circ$, $1.4^\circ$, $2.0^\circ$, and $3.0^\circ$. (b) Corresponding plots of $\sigma^y_{yy}(\omega;E_{\rm F})$ for the AB--BA variant. The missing component vanishes in the AB--AB/AB--BA variant due to the $C_{2x}$/$C_{2y}$ symmetry.}
    \label{fig:4-tdbg04-30}
\end{figure*}

We now examine how the SHG response evolves as a function of the twist angle $\theta$ in TDBG.  
Fig.~\ref{fig:4-tdbg04-30}(a) shows the band structures of AB--AB stacked TDBG together with the corresponding magnitude-phase density plots of $\sigma^x_{xx}(\omega;E_{\rm F})$ for several values of $\theta$.  
Because of the $C_{2x}$ symmetry of the AB--AB configuration, the $\sigma^y_{yy}$ component vanishes, and $\sigma^x_{xx}$ fully characterises the SHG response.

As $\theta$ is reduced from $3.0^\circ$ to $0.8^\circ$, the hyperbolic bands inherited from bilayer graphene are progressively reconstructed into moir\'{e} flat bands.
This evolution of the band structure is directly reflected in the SHG response.
As discussed in Sec.~\ref{subsec:resonant}, the formation of moir\'{e} flat bands promotes double-resonant optical processes by allowing the one-photon and two-photon resonance conditions to be simultaneously satisfied over extended regions of momentum space.
At the same time, the large joint density of states associated with the flat bands increases the number of available interband transitions within a narrow energy window.
As the twist angle decreases, these effects become increasingly pronounced, leading to a substantial enhancement of the SHG response.
In addition, the characteristic energy separations between neighbouring flat bands are reduced, causing the dominant optical transitions to shift toward lower frequencies, where the response is further amplified by the overall $1/\omega^2$ prefactor in Eq.~\eqref{eq:shgconducutivity}.

\begin{comment}
As $\theta$ is reduced from $3.0^\circ$ to $0.8^\circ$, the hyperbolic bands inherited from bilayer graphene are progressively reconstructed into moir\'{e} flat bands.  
%This reconstruction originates from the strengthening of the moir\'{e} interlayer coupling $U(\vec{r})$ in Eq.~\eqref{eq:tdbghamiltonian} as the twist angle decreases, which folds the spectrum into the moir\'{e} Brillouin zone and produces a dense set of relatively flat bands near the CNP.
This evolution of the band structure is directly reflected in the SHG response.  
As the twist angle is reduced and moir\'{e} flat bands develop, the joint density of states is substantially increased, allowing a large number of interband transitions to contribute within a narrow energy window.  
At the same time, the typical energy separations between neighbouring flat bands are reduced, so that the relevant optical gaps shift to lower energies.  
Both effects enhance the SHG response, since many $\alpha$ and $\beta$ channels are activated simultaneously and the resulting signal is further amplified by the overall $1/\omega^2$ prefactor in Eq.~\eqref{eq:shgconducutivity}.
\end{comment}

Fig.~\ref{fig:4-tdbg04-30}(b) shows the corresponding results for the AB--BA variant, where the finite component is $\sigma^y_{yy}$ due to the $C_{2y}$ symmetry.  
The overall evolution with twist angle follows the same qualitative trend as in the AB--AB case: as $\theta$ decreases, the band structure is reconstructed into moir\'{e} flat bands and the SHG response is strongly enhanced and develops increasingly rich frequency and Fermi-energy dependence.

\subsection{Effect of vertical bias voltage}\label{subsec:delta}
\begin{figure*}[t]
    \centering
    \includegraphics[width=17cm]{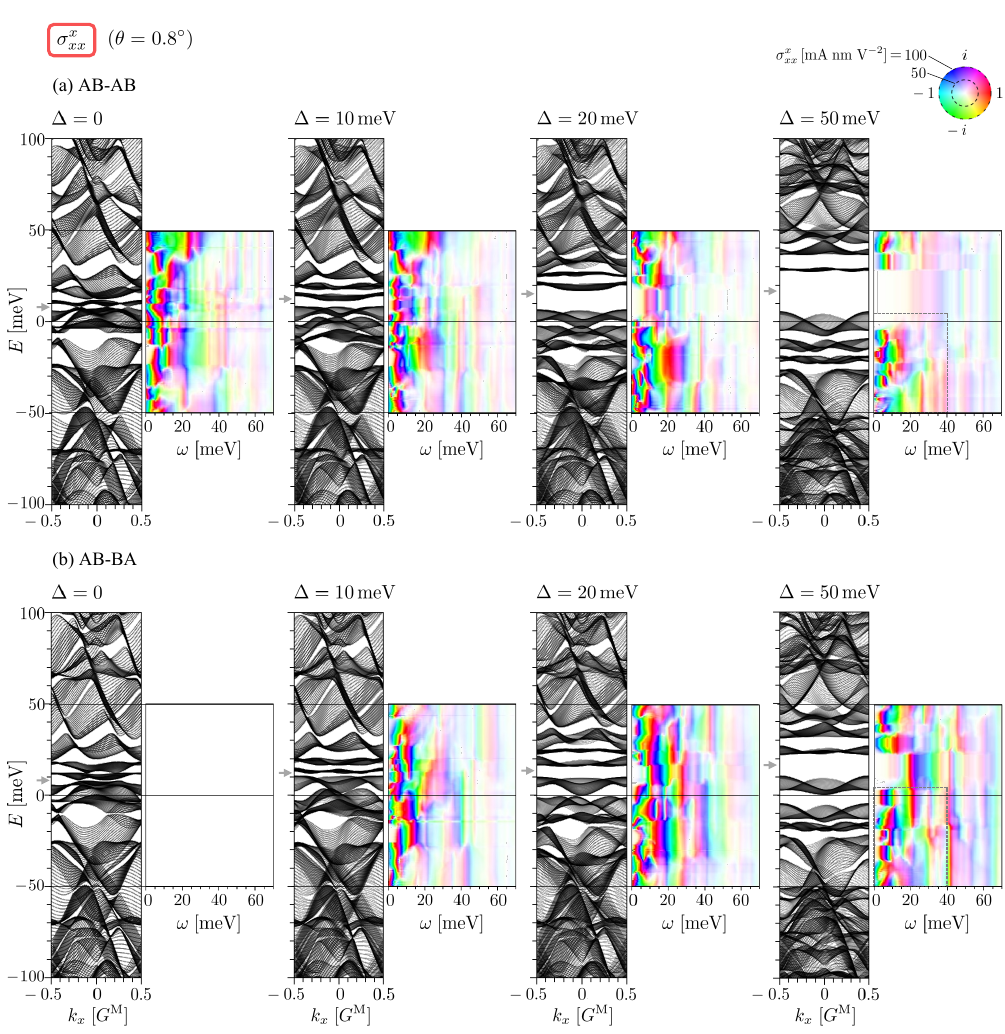}
    \caption{Band structures and $\sigma^x_{xx}(\omega;E_{\rm F})$ density plots of (a) AB--AB and (b) AB--BA stacked TDBG for several values of the vertical bias voltage $\Delta=$~0, 10, 20 and 50~meV at fixed twist angle $\theta=0.8^\circ$. The grey arrows indicate the corresponding charge-neutral gaps.}
    \label{fig:4-tdbg0800-50xxx}
\end{figure*}

\begin{figure*}[t]
    \centering
    \includegraphics[width=17cm]{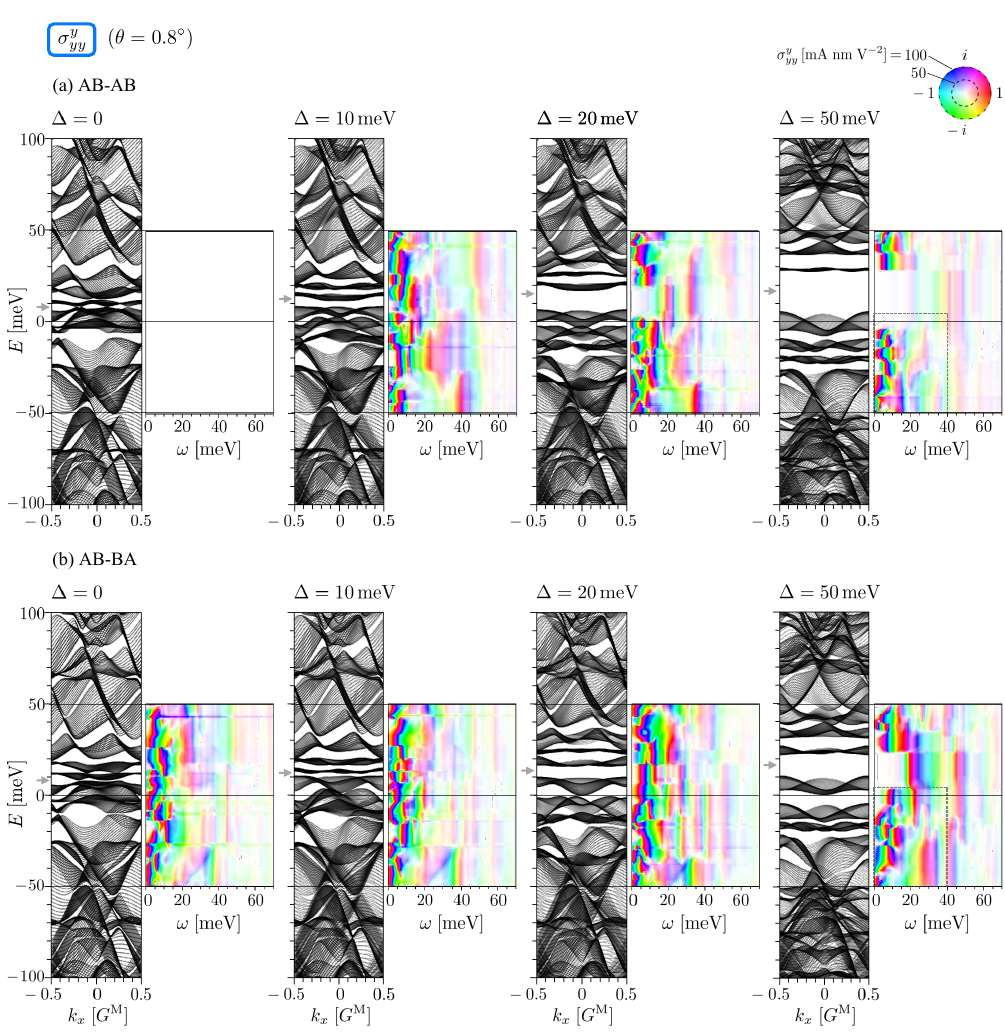}
    \caption{Band structures and $\sigma^y_{yy}(\omega;E_{\rm F})$ density plots of (a) AB--AB and (b) AB--BA stacked TDBG for several values of the vertical bias voltage $\Delta=$~0, 10, 20 and 50~meV at fixed twist angle $\theta=0.8^\circ$. The grey arrows indicate the corresponding charge-neutral gaps.}
    \label{fig:4-tdbg0800-50yyy}
\end{figure*}

We next turn to the effect of a vertical bias voltage $\Delta$ on the SHG response in TDBG at fixed twist angle $\theta=0.8^\circ$.  
The bias voltage affects the SHG signal in two complementary ways: it reshapes the low-energy moir\'{e} band structure by opening gaps at the CNP, and it alters the symmetry of the system, thereby controlling which components of the SHG tensor are allowed.

Fig.~\ref{fig:4-tdbg0800-50xxx}(a) shows the band structures and the corresponding $\sigma^x_{xx}(\omega;E_{\rm F})$ density plots of AB--AB stacking for several representative values of the vertical bias $\Delta=$~0, 10, 20 and 50~meV.
At small bias, we observe pronounced low-frequency features near the CNP, reflecting optical transitions between closely spaced moir\'{e} flat bands.
As $\Delta$ is increased, the CNP gap (grey arrows) widens, and these low-frequency contributions are progressively suppressed because the relevant interband transitions are pushed to higher energies.
By $\Delta=50$~meV, the enlarged CNP gap produces a clear insulating window around $E_{\rm F}=0$, indicated by a large white region.

At the same time, the vertical bias breaks the $C_{2y}$ symmetry and activates the $\sigma^x_{xx}$ component in the AB--BA variant.
This is illustrated in Fig.~\ref{fig:4-tdbg0800-50xxx}~(b), where the vanishing signal at $\Delta=0$ rapidly grows as a finite bias is applied.
With increasing $\Delta$, the overall magnitude increases, reaching a maximum around $\Delta\simeq20$~meV, and remains sizeable at larger bias.
%With increasing $\Delta$, the overall magnitude grows up to $\Delta\simeq 20$~meV and remains sizeable at larger bias.
Notably, at $\Delta=50$~meV the $\sigma^x_{xx}$ signal in AB--BA is substantially stronger than in AB--AB, reflecting energy spacings between neighbouring bands that are more favourable for double-resonant optical processes.
%Notably, at $\Delta=50$~meV the $\sigma^x_{xx}$ signal in AB--BA is substantially stronger than in AB--AB, reflecting the more favourable level spacing of the AB--BA bands for double-resonant processes.
The same qualitative trend is observed for $\sigma^y_{yy}$, with the roles of AB--AB and AB--BA interchanged, as shown in Fig.~\ref{fig:4-tdbg0800-50yyy}.

\subsection{Comparison between AB--AB and AB--BA stacked TDBG}\label{subsec:comparison}
\begin{comment}
    \begin{figure}[t]
    \centering
    \includegraphics[width=8.5cm]{figures/4_0850_zoomin_v1.0.pdf}
    \caption{[(a) and (c)] Close-up view of the $\sigma^x_{xx}$ and $\sigma^y_{yy}$ responses in AB--AB TDBG shown in Figs.~\ref{fig:4-tdbg0800-50xxx} and \ref{fig:4-tdbg0800-50yyy}. [(b) and (d)] Corresponding colour-density plots for the AB--BA variant. We have labelled regions where a systematic $\pi$ phase shift between the two stacking configurations can be seen.}
    \label{fig:4-tdbg0850_zoom}
\end{figure}

\end{comment}
\begin{figure}[t]
    \centering
    \includegraphics[width=8.5cm]{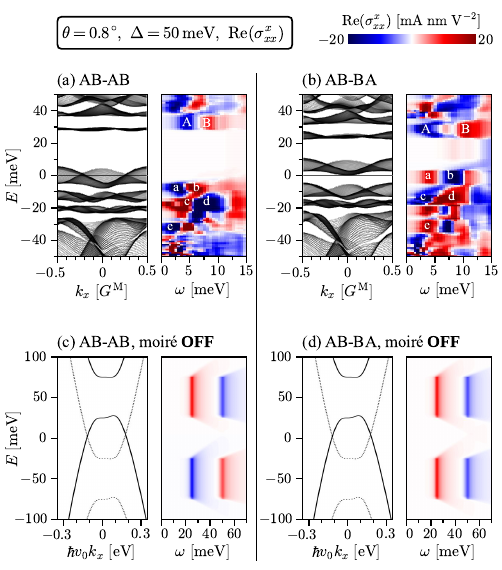}
    \caption{[(a) and (b)] Close-up view of the real parts of $\sigma^x_{xx}$ response in AB--AB and AB--BA TDBG shown in Figs.~\ref{fig:4-tdbg0800-50xxx} and \ref{fig:4-tdbg0800-50yyy}. [(c) and (d)] Corresponding band and density plots for the systems with the moir\'{e} interaction turned off.}
    \label{fig:4-tdbg0850_zoom}
\end{figure}

We now compare the SHG responses of the AB--AB and AB--BA variants and highlight a systematic phase relationship between them.
Figure~\ref{fig:4-tdbg0850_zoom}(a) and (b) show the low-frequency region of $\sigma^x_{xx}(\omega,E_{\rm F})$ for both stackings at $\theta=0.8^\circ$ and $\Delta=50$~meV (the corresponding full spectra are shown in the rightmost panels of Fig.~\ref{fig:4-tdbg0800-50xxx}), where the contrast between the two configurations is most pronounced.
The colour map represents the real part of $\sigma^x_{xx}$ (red and blue indicating positive and negative values, respectively) in order to emphasise the phase relationship between the two stackings.
Throughout the valence-band region, we find an approximate $\pi$ phase shift between the two variants at corresponding positions in the $(\omega,E_{\rm F})$ plane, labelled a--e for clarity.
By contrast, in the conduction-band region (A and B), the phases are approximately aligned.
The same phase relationship is also observed in the imaginary part of $\sigma^x_{xx}$.

The origin of this phase relationship can be understood by first considering the SHG response of the uncoupled bilayers.
Figures~\ref{fig:4-tdbg0850_zoom}(c) and (d) show the corresponding spectra with the moir\'e interlayer coupling switched off, i.e., two independent bilayer graphenes rotated by $\pm\theta/2$ [Fig.~\ref{fig:1-geometry}].
The lower and upper bilayers have band gaps centred at $E=\pm50$~meV, respectively, owing to the interlayer potential $\Delta=50$~meV.
Within these gaps, the SHG features at $\omega\approx25$ and $50$~meV correspond to the $\beta_{23}$ and $\alpha_{23}$ resonances of bilayer graphene discussed in Fig.~\ref{fig:3-blg}.

In the AB--AB configuration, the SHG signals from the two bilayers have opposite signs.
This originates from the fact that, in the absence of rotation, $\sigma^x_{xx}$ vanishes by symmetry, while a small rotation by $\pm\theta/2$ generates SHG signals with opposite phases in the two bilayers.
In the AB--BA configuration, however, the additional $180^\circ$ rotation of the upper bilayer reverses the sign of its SHG response, causing the signals from the lower and upper bilayers to become phase aligned.
Consequently, when the AB--AB and AB--BA configurations are compared, the SHG signal associated with the upper bilayer ($E\approx-50$~meV) changes sign, whereas that associated with the lower bilayer ($E\approx+50$~meV) remains nearly unchanged.

Remarkably, this simple phase relationship survives the strong moir\'e hybridisation.
As seen in Figs.~\ref{fig:4-tdbg0850_zoom}(a) and (b), the low-frequency SHG response of the fully coupled TDBG retains essentially the same relative phase structure despite the substantial reconstruction and mixing of the electronic bands.
An analogous sign reversal between the AB--AB and AB--BA stackings was previously identified in the shift current of TDBG~\cite{JoyaKawakamiKoshino2025}, indicating that this phase relation is a robust consequence of the stacking geometry.

\section{Conclusion}\label{sec:conclusion}

In this paper, we have studied second-harmonic generation (SHG) in non-moir\'{e} AB-stacked bilayer graphene and twisted double bilayer graphene (TDBG), with the aim of clarifying how moir\'{e} band reconstruction reshapes the second-order optical response.
A direct comparison between the non-moir\'{e} bilayer system and TDBG shows that the moir\'{e} system generically exhibits a much stronger SHG signal, owing to the formation of moir\'{e} flat bands and the associated enhancement of both the joint density of states and resonant optical processes.

We then presented a systematic analysis of the SHG spectrum in TDBG as a function of twist angle, vertical bias voltage, Fermi energy, and stacking configuration.
We showed that the formation of moir\'{e} flat bands promotes double-resonant optical processes, in which the one-photon and two-photon resonance conditions are simultaneously satisfied over extended regions of the moir\'{e} Brillouin zone.

As the twist angle is reduced and the bands become flatter, these double-resonant channels become increasingly abundant, resulting in a strong enhancement of the SHG response.
We further found that the detailed structure of the SHG spectrum evolves sensitively with the Fermi energy and vertical bias.
As the Fermi level is varied, the dominant resonant transitions change between different sets of moir\'{e} bands, producing abrupt modifications of both the magnitude and phase of the response.
More broadly, these results demonstrate how moir\'{e} reconstruction effectively reproduces the resonance enhancement familiar from systems with discrete energy levels, providing a general mechanism for strong nonlinear optical responses in flat-band moir\'{e} materials.

Finally, by comparing AB--AB and AB--BA TDBG, we found a systematic relative $\pi$ phase shift between the SHG signals below charge neutrality.
This phase offset originates from the $180^\circ$ rotation of the second bilayer in the AB--BA structure relative to AB--AB, which reverses the in-plane polarisation and the effective direction of the response.
Notably, this stacking-induced relative phase, which is already present in the uncoupled TDBG model, survives the strong moir\'{e} band reconstruction.

\begin{acknowledgements}
The authors acknowledge fruitful discussions with Vladimir Fal'ko, Takahiro Morimoto, and Kenichi Asano.
This work was supported by JSPS KAKENHI Grants No. JP20K14415, No. JP20H01840, No. JP20H00127, No. JP21H05236, No. JP21H05232, JP24K06921, by JST CREST Grant No. JPMJCR20T3, and by JST SPRING, Grant No. JPMJSP2138, Japan.
\end{acknowledgements}
%%%%%%%%%%%%%%%%%%%%%%%%%%%%%%%%%%%%%%%%%%%%%%%%%%%%%%%%%%%%%%%%%%%%%%%%%%%%%%%%%%%%%%%%%%%%%%%%
%%%%%%%%%%%%%%%%%%%%%%%%%%%%%%%%%%%%%%%%%%%%%%%%%%%%%%%%%%%%%%%%%%%%%%%%%%%%%%%%%%%%%%%%%%%%%%%%

%%%%%%%%%%%%%%%%%%%%%%%%%%%%%%%%%%%%%%%%%%%%%%%%%%%%%%%%%%%%%%%%%%%%%%%%%%%%%%%%%%%%%%%%%%%%%%%%%%%%%%%%%%%%%%%%%%%%%%
\appendix
\section{Broadening parameter in the SHG conductivity tensor}\label{appx:comparison}
\begin{figure}[t!]
    \centering
    \includegraphics[width=8.5cm]{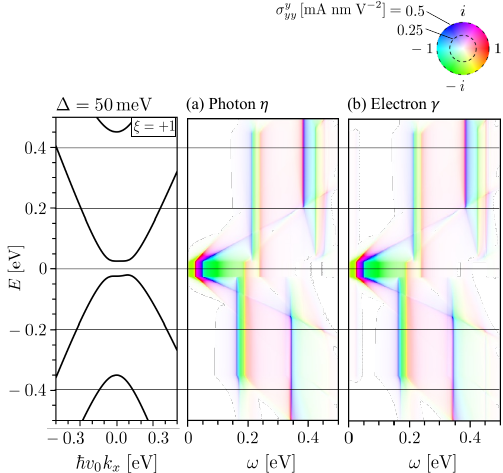}
    \caption{Band structure of AB-stacked bilayer graphene at the $K_+$ valley and the $\sigma^y_{yy}$ response using (a) the photon broadening $\eta$ and (b) the electron broadening $\gamma$.}
    \label{fig:A-matsubarakeldysh}
\end{figure}

In this Appendix, we discuss how the broadening parameter should enter in the explicit forms of the SHG conductivity tensor in the velocity gauge, which is known to be a delicate issue in the community~\cite{PassosVentura2018,ParkerMorimoto2019,SarsfieldJoyaFalko2026}.

In the present work, we have derived $\sigma^\mu_{\alpha\alpha}$ in Eq.~\eqref{eq:shgconducutivity} using the temperature Green's function, and analytically continuing the Matsubara frequencies $i\omega$ to real frequencies $\hbar\omega$ by replacing $i\omega\to\hbar\omega+i\eta$ and $2i\omega\to2\hbar\omega+2i\eta$.
The second replacement assumes that the phenomenological broadening originates from the adiabatic switch on of the external field.
In other words, $\eta$ is the photon broadening and the external electric field is written as
\begin{equation}
    E(t)=\int_{-\infty}^\infty\frac{d\omega}{2\pi}\,E(\omega)\,e^{i\omega t+\eta t},
\end{equation}
such that $E(t)$ vanishes in the infinite past.

\begin{widetext}
Although we will not step in to the technical details, using the temperature Green's function, the SHG conductivity under interest can be given as~\cite{ParkerMorimoto2019}
\begin{align}
    \sigma^\mu_{\alpha\alpha}(\omega) &= \frac{e^3}{2\pi^2\omega^2}\frac{1}{\beta}\sum_n\int d^2k\sum_{a,b,c}\frac{v^\mu_{ac}v^\beta_{cb}v^\alpha_{ba}}{(i\varepsilon_n + 2i\omega - \varepsilon_c)(i\varepsilon_n + i\omega - \varepsilon_b)(i\varepsilon_n - \varepsilon_a)} \nonumber
    \\
    &= \frac{e^3}{2\pi^2\omega^2}\int d^2k\sum_{a,b,c}\frac{v^\mu_{ac}v^\beta_{cb}v^\alpha_{ba}}{2i\omega - \varepsilon_{ca}}\left(\frac{f_{ab}}{i\omega - \varepsilon_{ba}}-\frac{f_{bc}}{i\omega - \varepsilon_{cb}}\right),
\end{align}
where $\beta=(k_{\rm B}T)^{-1}$ is the inverse temperature, $i\varepsilon_n = 2\pi(n+1)k_{\rm B}T$ and $i\omega$ are the fermionic and bosonic Matsubara frequencies, respectively.
As a result, we obtain the following expression for $\sigma^\mu_{\alpha\alpha}$,
\begin{align}
    \sigma^\mu_{\alpha\alpha}(\omega) &= \frac{e^3}{2\pi^2\omega^2}\int d^2k \sum_{a,b,c}\frac{v^\mu_{ac}v^\alpha_{cb}v^\alpha_{ba}}{2\hbar\omega-\varepsilon_{ca}+2i\eta}\left(\frac{f_{ab}}{\hbar\omega-\varepsilon_{ba}+i\eta}-\frac{f_{bc}}{\hbar\omega-\varepsilon_{cb}+i\eta}\right) \nonumber
    \\
    &= \frac{e^3}{2\pi^2\omega^2}\int d^2k\sum_{a,b,c}\frac{v^\mu_{ac}v^\alpha_{cb}v^\alpha_{ba}}{\varepsilon_{ba}-\varepsilon_{cb}}\left(\frac{f_{ab}}{\hbar\omega-\varepsilon_{ba}+i\eta}+\frac{f_{bc}}{\hbar\omega-\varepsilon_{cb}+i\eta}-\frac{2f_{ac}}{2\hbar\omega-\varepsilon_{ca}+2i\eta}\right),
    \label{eq:matsubara_shg}
\end{align}
arriving at Eq.~\eqref{eq:shgconducutivity}, where the broadening of the 2-photon resonance is double that of the 1-photon resonances.
We note that the factor of 2 is crucial in the last line of the expression; if the final $2i\eta$ is replaced by a simple $i\eta$, the expression begins to violate conservation laws which lead to spurious divergences once double resonance is achieved.

On the other hand, one can take a different approach and implement the phenomenological broadening $\gamma$ of the electrons, instead of the photons $\eta$.
This can be achieved through a replacement of the form, $i\omega\to\hbar\omega+i\gamma$ and $2i\omega\to2\hbar\omega+i\gamma$~\footnote{This result reproduces the expression obtained by using the Keldysh formulation of nonequilibrium Green's functions, as done in Ref.~\cite{SarsfieldJoyaFalko2026}. Namely, replace the factors of $\eta_a+\eta_b$ with $\gamma$.}, thus, the SHG conductivity reads
\begin{align}
    \sigma^\mu_{\alpha\alpha}(\omega) &= \frac{e^3}{2\pi^2\omega^2}\int d^2k \sum_{a,b,c}\frac{v^\mu_{ac}v^\alpha_{cb}v^\alpha_{ba}}{2\hbar\omega-\varepsilon_{ca}+i\gamma}\left(\frac{f_{ab}}{\hbar\omega-\varepsilon_{ba}+i\gamma}-\frac{f_{bc}}{\hbar\omega-\varepsilon_{cb}+i\gamma}\right) \nonumber
    \\
    &= \frac{e^3}{2\pi^2\omega^2}\int d^2k \sum_{a,b,c}v^\mu_{ac}v^\alpha_{cb}v^\alpha_{ba}\left[\frac{1}{\varepsilon_{ba}-\varepsilon_{cb}-i\gamma}\frac{f_{ab}}{\hbar\omega-\varepsilon_{ba}+i\gamma}+\frac{1}{\varepsilon_{ba}-\varepsilon_{cb}+i\gamma}\frac{f_{bc}}{\hbar\omega-\varepsilon_{cb}+i\gamma}\right. \nonumber
    \\
    &\hspace{4.8cm}\left.-2\left(\frac{f_{ab}}{\varepsilon_{ba}-\varepsilon_{cb}-i\gamma}+\frac{f_{bc}}{\varepsilon_{ba}-\varepsilon_{cb}+i\gamma}\right)\frac{1}{2\hbar\omega-\varepsilon_{ca}+i\gamma}\right],
    \label{eq:keldysh_shg}
\end{align}
where we now have a slightly more complicated resonance structure compared to Eq.~\eqref{eq:matsubara_shg}.

Now that we have two expressions for $\sigma^\mu_{\alpha\alpha}$, we can compare their numerical behaviours.
The SHG signal in AB-stacked bilayer graphene has been calculated using the two expressions, where we show the numerical results in Fig.~\ref{fig:A-matsubarakeldysh}.
Fig.~\ref{fig:A-matsubarakeldysh} (a) shows the SHG spectrum with electron broadening $\gamma=1$~meV while (b) shows that for the photon broadening $\eta=1$~meV.
We can see that the results are practically identical, both in terms of phase and magnitude, and the physics is relatively indifferent to which broadening one adopts, as long as $\gamma$ and $\eta$ are chosen to be small enough compared to the typical energy scale~\cite{PassosVentura2018}, such as the gap size or dispersion.
In our current studies, we have opted for $\eta$ since the analytical form of $\sigma^\mu_{\alpha\alpha}$ was simpler, thus, providing a clearer physical picture.
We finally note that the above discussion is completely independent of the dimensionality, which was 2 in our case to retain consistency with the main text.

\section{Derivation of the double-resonant signal}\label{appx:dr_analysis}

In the present Appendix, we give a brief derivation of the double-resonant signal given in Eq.~\eqref{eq:shg_dr}.
We begin by considering a specific triplet of states $(a,b,c)$ with energies $(\varepsilon_a,\varepsilon_b,\varepsilon_c)=(-\varepsilon,\delta\varepsilon,\varepsilon)$.
These states are very close to being equidistant, hence, close to double resonance.
In the limit of $\delta\varepsilon\to0$, the contribution to the SHG signal from the transition between $(a,b,c)$ can be calculated from  Eq.~\eqref{eq:shgconducutivity} as
\begin{align}
    \sigma^\mu_{\alpha\alpha}(\omega) &\sim \frac{e^3}{2\pi^2\omega^2}\int d^2k\;\frac{v^\mu_{ac}v^\alpha_{cb}v^\alpha_{ba}}{2\delta\varepsilon}\left(\frac{f_{ab}}{\hbar\omega-\varepsilon-\delta\varepsilon+i\eta}+\frac{f_{bc}}{\hbar\omega-\varepsilon+\delta\varepsilon+i\eta}-\frac{2f_{ac}}{2\hbar\omega-2\varepsilon+2i\eta}\right) \nonumber
    \\
    &\sim \frac{e^3}{2\pi^2\omega^2}\int d^2k\;\frac{v^\mu_{ac}v^\alpha_{cb}v^\alpha_{ba}}{2\delta\varepsilon}\left(\frac{f_{ab} - f_{bc}}{(\hbar\omega-\varepsilon+i\eta)^2}\delta\varepsilon+\frac{f_{ac}}{(\hbar\omega-\varepsilon+i\eta)^3}\delta\varepsilon^2 + \mathcal{O}(\delta\varepsilon^3)\right) \nonumber
    \\
    &\sim \frac{e^3}{4\pi^2\omega^2}\int d^2k\,\left(f_{ab} - f_{bc}\right)\frac{v^\mu_{ac}v^\alpha_{cb}v^\alpha_{ba}}{(\hbar\omega-\varepsilon+i\eta)^2}+\mathcal{O}(\delta\varepsilon),
    \label{eq:matsubara_shg_dr}
\end{align}
where we arrive at Eq~\eqref{eq:shg_dr}.
\begin{comment}
If the electron broadening $\gamma$ were to be adopted instead, one would arrive at a slightly modified expression~\footnote{This expression differs from that given in Eq.~(B24) in Ref.~\cite{SarsfieldJoyaFalko2025} due to the difference in the energy dependence of the broadening. Here, we adopted the same broadening $\gamma/2$ for all states, while in Ref~\cite{SarsfieldJoyaFalko2025}, they set $\eta_b=0$ for the state near the Fermi energy and $\eta_a=\eta$ for all other states.}
\begin{align}
    \sigma^\mu_{\alpha\alpha}(\omega) \sim \frac{e^3}{4\pi^2\omega^2}\int d^2k\,\frac{\left(f_{ab} - f_{bc}\right)\,v^\mu_{ac}v^\alpha_{cb}v^\alpha_{ba}}{(\hbar\omega-\varepsilon+i\gamma)(\hbar\omega-\varepsilon+2i\gamma)}+\mathcal{O}(\delta\varepsilon).
    \label{eq:keldysh_shg_dr}
\end{align}
\end{comment}
\end{widetext}

\bibliography{reference}
\end{document}